# Nucleation of titanium nanoparticles in an oxygen-starved environment, I: Experiments

Rickard Gunnarsson[1], Nils Brenning[1, 2]*, Robert Deric Boyd[1], and Ulf Helmersson[1]

[1] IFM-Material Science, Linköping University, 581 83 Linköping, Sweden
[2] KTH Royal Institute of Technology, School of EECS, Department of Space and Plasma Physics, SE-100 44, Stockholm, Sweden
*corresponding author (e-mail: nils.brenning@ee.kth.se)

**Abstract**: A constant supply of oxygen has been assumed to be necessary for the growth of titanium nanoparticles by sputtering. This oxygen supply can arise from a high background pressure in the vacuum system or from a purposely supplied gas. The supply of oxygen makes it difficult to grow metallic nanoparticles of titanium and can cause process problems by reacting with the target. We here report that growth of titanium nanoparticles in the metallic hexagonal titanium (αTi) phase is possible using a pulsed hollow cathode sputter plasma and adding a high partial pressure of helium to the process instead of trace amounts of oxygen. The helium cools the process gas in which the nanoparticles nucleate. This is important both for the first dimer formation and the continued growth to a thermodynamically stable size. The parameter region, inside which the synthesis of nanoparticles is possible, is mapped out experimentally and the theory of the physical processes behind this process window is outlined. A pressure limit below which no nanoparticles were produced was found at 200 Pa, and could be attributed to a low dimer formation rate, mainly caused by a more rapid dilution of the growth material. Nanoparticle production also disappeared at argon gas flows above 25 sccm. In this case the main reason was identified as a gas temperature increase within the nucleation zone, giving a too high evaporation rate from nanoparticles (clusters) in the stage of growth from dimers to stable nuclei. These two mechanisms are in depth explored in a companion paper. A process stability limit was also found at low argon gas partial pressures, and could be attributed to a transition from a hollow cathode discharge to a glow discharge.

## 1. Introduction

This is an experimental study of the nucleation of titanium nanoparticles in an oxygen-starved environment, and it goes together with a theoretical companion paper [1]. The growth of nanoparticles in the gas phase from vapor created by sputtering has several advantages over other synthesis methods. To name a few, it allows for well dispersed particles on surfaces, a wide choice of different materials [2] and the benefits of being a continuous process. Other advantages are the high purity compared to liquid phase synthesis methods, where the nanoparticles easily get contaminated by trace elements present in the liquid media [3].

Contamination, however, can be a problem also in gas phase synthesis, particularly for highly reactive materials. We have previously reported that for the case of titanium, residual gases in the vacuum system can significantly contaminate the particles [4] [5] favoring the formation of titanium(II) oxide rather than metallic titanium nanoparticles. It was also found if the residual gas content was reduced no formation of nanoparticles occurred. That means metallic nanoparticles could not be obtained. The same behavior has been demonstrated by several other researchers [5] [6] [7] [8]. This behavior has been attributed to a higher binding energy between the metal and oxygen atoms compared to metal and metal atoms, causing the formation of dimers with a higher stability [8]. Titanium is not the only element that is affected by residual gases in the formation of nanoparticles. Similar results have been observed for silver [9], vanadium [10], tungsten [11] and cobalt [8]. The addition of oxygen into a sputtering process can be used to stimulate nucleation but can, besides unwanted nanoparticle oxidation, introduce problems such as cathode poisoning (resulting in decreased sputter yields), arcing, and process hysteresis [12]. It has also been shown that only certain discharge conditions give a stable productivity of particles when oxygen is added [7].

It is thus clear that a different way of growing nanoparticles is of high interest, which is the focus of the present paper. By introducing a high partial pressure of helium and combining it with a pulsed hollow cathode discharge, nanoparticles are synthesized in an ultrahigh vacuum system without the need of adding oxygen. The effect of the helium is to decrease of the gas temperature within the region where the ejected titanium vapors have its highest density.

## 2. Experimental setup

The nanoparticle deposition source, schematically drawn in figure 1, consists of a hollow cathode within which the sputtering occurs. The material gets extracted into a growth zone, where the first dimers form and grow to a stable size $r^*$. These two steps, together, constitute the nucleation process. Further growth of the nanoparticles occurs between the cathode and the anode. The nanoparticles get transported to the substrates by the gas flow, and by the electric field from the substrate bias. The nanoparticle deposition source was pumped to ultrahigh vacuum conditions in the low $10^{-7}$ Pa range. The substrate table was positioned in a high vacuum system, where the base pressure was in a mid $10^{-5}$ Pa range. The two systems were separated by a gate valve. During the nanoparticle deposition, this gate valve was opened while the gate valve to the nanoparticle source turbopump was closed. To suppress diffusion from the high vacuum system to the ultrahigh vacuum system, argon gas was flowing while the gate valve between the systems was open. The argon gas (99.9997 % purity) passed through a gas purifier (Ultra pure) before entering the chamber. The helium gas (99.99990 % purity) passed through a gas dryer (Mini Dryer XL) and was injected under copper gaskets that seal the flanges, in order to increase its dispersion in the chamber. It is also possible to flow oxygen through the helium gas inlet. To further reduce the contaminants, the gas supply lines were differentially pumped prior to the deposition by a parallel connection to the high vacuum system. The gas-flows were controlled using mass-flow regulators (Mass-Flo Controller, MKS) with an upper flow limit of 500 sccm for the Ar and He gases and an upper limit to the oxygen flow of 20 sccm. To further increase the precision in the oxygen flow, the oxygen was supplied from an argon-oxygen gas mixture (0.5 mol% oxygen). The process pressure was measured by a capacitance manometer and automatically regulated to a set point value by a throttle valve. This allowed for the process pressure to be set independently of the argon gas flow. The hollow cathode consists of a 55 mm long titanium tube of 99.6% purity with an inner diameter of 5 mm. It was clamped in a water cooled copper block isolated from the plasma using a fiberglass weave. A ring-shaped anode with a diameter of 34 mm was positioned 64 mm from the hollow cathode exit and was maintained at a potential of 43 V during operation. The diameter of the growth chamber was 98 mm. The substrate table was positioned 294 mm from the hollow cathode and the 10x10 mm$^2$ gold coated silicon substrates had a bias potential in the range 150-200 V. The design of the substrate table made it possible to transfer the substrates to a window, for optical inspection, without breaking vacuum. By wrapping the chamber with copper wires and leading them to a bath of liquid nitrogen, the walls of the nanoparticle deposition source could be cooled down to 225 K. By instead wrapping it with resistive heating bands, they could be heated up above room temperature.

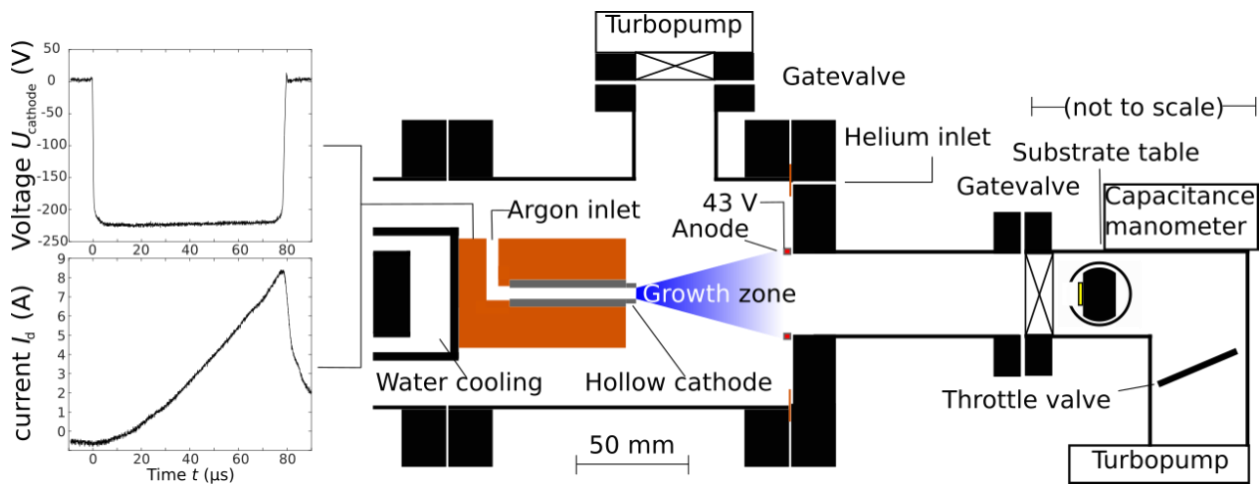

*Figure 1 Experimental setup. The UHV part, containing the hollow cathode, is separated from the part containing the substrate table by a gate valve. The argon gas flow is passed through the hollow cathode into the growth chamber. The helium gas is injected under a copper gasket to increase its spread in the chamber. The process pressure is set by the throttle valve. The nanoparticles are guided by the gas flow and substrate bias towards the substrate table. The gas flow also suppresses backflow of contaminants from the substrate table. A typical discharge current and voltage pulse is shown. The figure is to scale except for the region where the substrate table is depicted.*

The power to the cathode was supplied by an in-house built pulsing unit connected to a DC power supply (MDX 1K). The DC power supply was operated in constant current mode with a set point value of 0.52 A. The pulsing frequency was set to 1500 Hz with a pulse width of 80 μs which resulted in peak voltages of -217 to -255 V and peak currents of 7 to 11.5 A, depending on the pressure. This resulted in average powers of around 100 W. The nanoparticle size was evaluated by measuring their diameter from scanning electron microscopy (SEM) images taken by a LEO 1550 Gemini microscope. Transmission electron microscopy (TEM) was performed using a FEI Technai G[2] TF 20 UT microscope and x-ray diffractometry (XRD) was performed on an PANalytical empyrean x-ray diffractometer operated at in grazing incidence mode.

## 3. Results

For the experiments in this work, we keep the pulsing parameters constant and vary four external parameters: total pressure $p$, argon gas flow $Q_{\text{Ar}}$, helium gas flow $Q_{\text{He}}$ and wall temperature $T_{\text{wall}}$. We also add a flow of oxygen $Q_{\text{O}_2}$ to the helium gas inlet for selected experiments. The existence of nanoparticles independent of size was first evaluated in order to see under which parameters nucleation (dimer formation followed by growth to stable nanoparticle size $r^*$) occurs. We have previously found [5] that $p$ and $Q_{\text{Ar}}$ are two determining parameters for nucleation in high vacuum. Analyzing nanoparticle production in this 2-d parameter space is therefore informative. Such ($p$, $Q_{\text{Ar}}$) surveys will be a main tool also in this work.

### 3.1. A process instability

It was first attempted to obtain nanoparticles in our ultrahigh vacuum system (*i.e.,* a system with very little water vapor contamination) with only argon. These attempts were unsuccessful and resulted in no production of particles. The addition of helium made it possible to obtain nanoparticles, but too much helium introduced a discharge instability which limited the size of the useful parameter range. Helium therefore opened a process window, which we here only had resources to investigate for one single He flow, where we arbitrarily have chosen 55 sccm. The limited aim here is to understand the physics that constrain the process window. Optimization within the 3-d space ($p$, $Q_{\text{Ar}}$, $Q_{He}$) remains to be done.

The process instability manifested itself in an increased peak height of the discharge current which caused a behavior similar to arcing. The process could not be run under these conditions without causing damage to the experimental setup. The data points in a ($p$, $Q_{\text{Ar}}$) survey where this process instability occurred is marked by circles in figure 2. The limit was found to fit well with the gas flow that would maintain a constant argon gas partial pressure when assuming a complete mixing of the two process gases (blue dot-dashed line). The equation for this line and the reason for the process instability will be further discussed in the theory section. This process instability gives a lower argon gas flow limit to the useful process window in the ($p$, $Q_{\text{Ar}}$) parameter space.

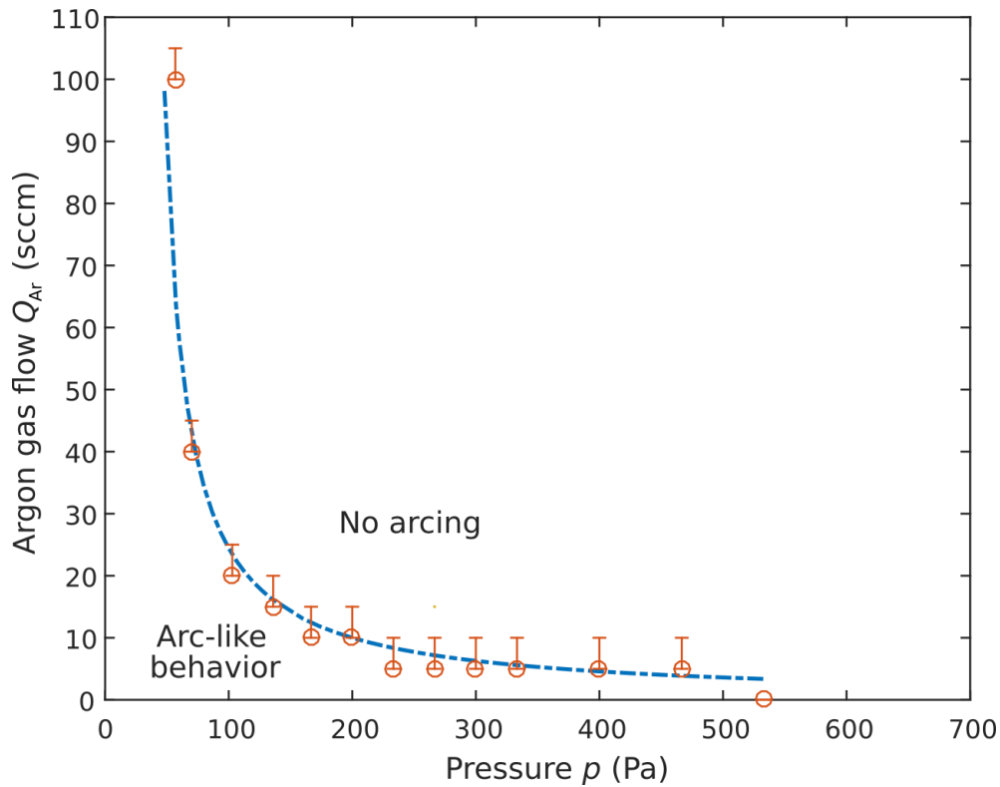

Figure 2 A (p, $Q_{Ar}$) survey of the process instability at a constant helium gas flow of 55 sccm. At pressures higher than 533 Pa the discharge was stable in pure helium, also at zero argon gas flow. Below that pressure circles mark the combinations (p, $Q_{Ar}$) at which there was a discharge instability, an arc-like behavior. The blue dot-dashed line is a theoretical curve with a good fit to the experimental data: the combinations (p, $Q_{Ar}$) at which there is a constant argon gas partial pressure of 28.27 Pa in the growth zone.

At higher pressures than 533 Pa it was possible to run the process with only helium supplied; however, this was not a useful parameter range since no nanoparticles were found without an argon gas flow.

### 3.2. Limits in nucleation: are impurities involved?

The process instability limits the available discharge parameter range for nanoparticle production. There is also an upper argon gas flow limit where no nanoparticles were observed by ocular inspection after 10 minutes of deposition, herein referred to as the "$Q_{Ar}$ limit". It is marked in figure 3 by a horizontal dashed black line fitted to the data points. The $Q_{Ar}$ limit varied between experiments as illustrated by the error bars, but always averaged at around $Q_{Ar}$= 20-35 sccm for the full pressure range investigated. There is no clear pressure dependence of the $Q_{Ar}$ limit. Below 200 Pa no nanoparticles were found by ocular inspection at any value of $Q_{Ar}$. We call this the "$p$ limit". It is marked by a vertical red dotted line. The reasons for these two limits are discussed in the companion paper [1].

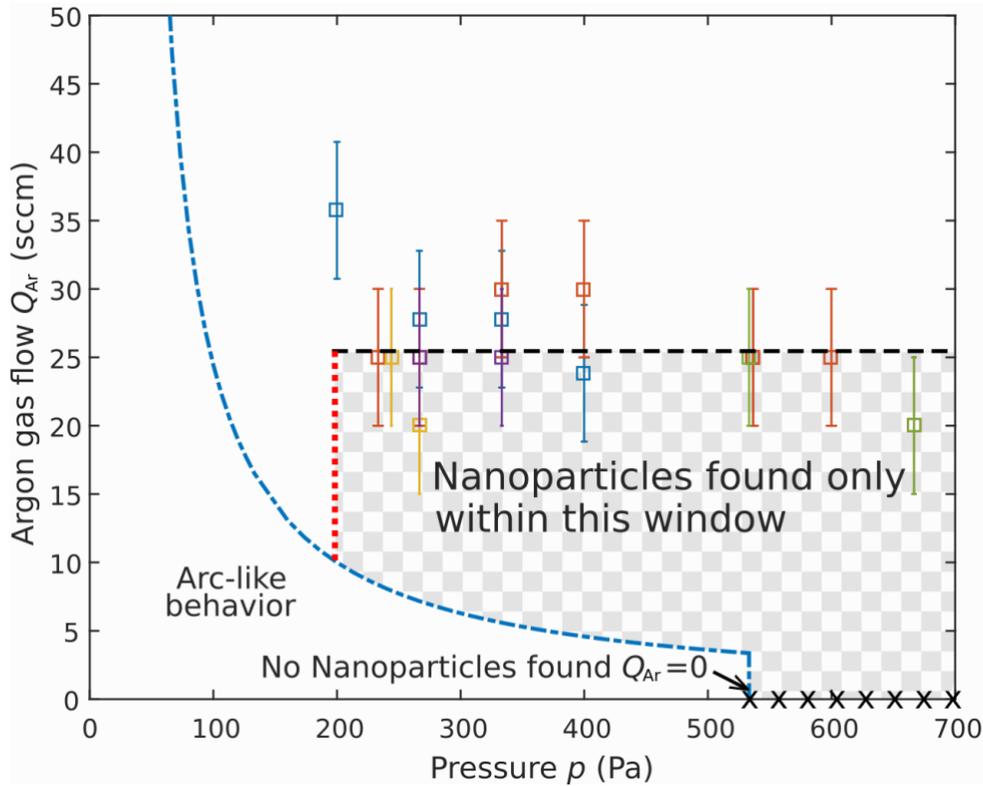

*Figure 3 The process window within which nanoparticles are found in a (p, $Q_{Ar}$) survey, with a constant helium gas flow of $Q_{He} = 55$ sccm. The data points for disappearance of nanoparticles is marked by squares. The error bars denote the uncertainty of the measurement method. The horizontal dashed black line represents the $Q_{Ar}$ limit. Above it, no nanoparticles were on average found after 10 minutes of deposition. Below the blue dot-dashed line, the process became unstable. The vertical red dotted line represents the p limit, to the left of which no nanoparticles were found. No nanoparticles where found at an argon gas flow of 0 sccm, which is marked by crosses. The process window where nanoparticles were found is within the checkered area.*

We now want to test the hypothesis that impurities such as oxygen or water vapor assist in the nucleation and/or growth processes. Increased impurity levels should in that case increase the size of the process window. The hypothesis is first tested by manipulating the degassing inside the vacuum system by heating or cooling the vacuum chamber wall. The helium flow $Q_{He}$ was kept constant at 55 sccm. The gas density in the growth zone was, to a first approximation, kept constant during these experiments by varying the pressure with the throttle valve to compensate for the temperature change, keeping $p/T_{wall}$ constant[1]. The pressure was thus increased from 200 Pa to 377 Pa when the wall temperature was increased from 225 to 425 K. To find the $Q_{Ar}$ limit for the appearance of particles on the substrate, the flow was decreased with steps of 5 sccm. As can be seen in figure 4 (a) it is first clear that the actual limit varies between experiments run at the same parameters. If instead the trends are looked at, it can be seen that there is no clear trend between $T_{wall}$ = 225 K and 350 K. As the temperature is increased above 350 K there is a reproducible increase in the $Q_{Ar}$ limit. This increase is consistent with the higher degassing rate of adsorbed species on the vacuum chamber wall. An approximation of how the wall temperature influences the vapor pressure of adsorbed water molecules is plotted in the figure as a black line. The approximations made for this plot will be covered in the discussion section.

---

[1] From the gas law in the form $p \propto n_g/T_g$ follows that this method only keeps the gas density constant at locations where the gas has a temperature that is proportional to $T_{wall}$. Inside, and close to, the hollow cathode, this method therefore would not keep the gas density constant.

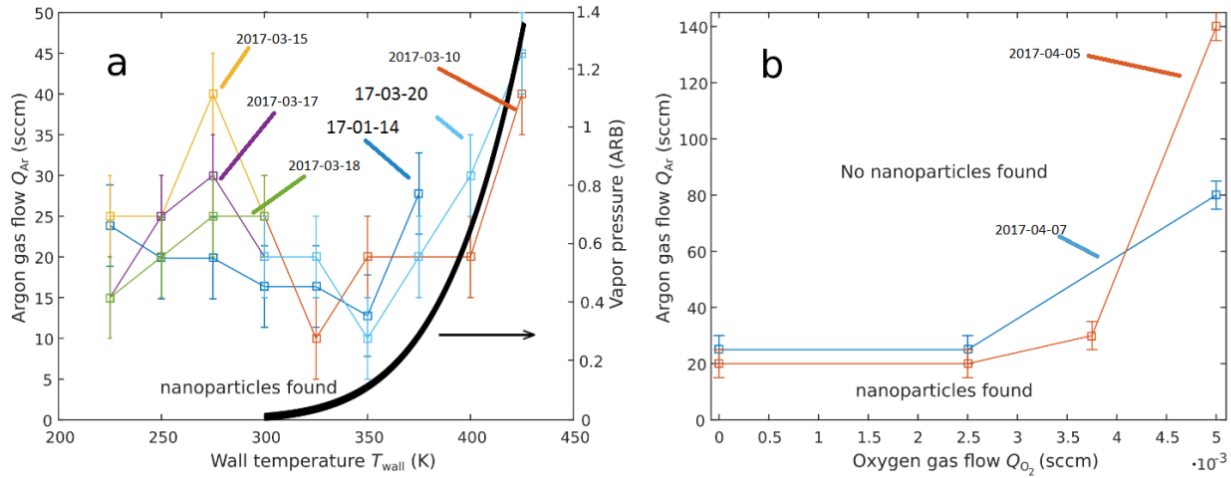

Figure 4 (a) The $Q_{Ar}$ limit as a function of chamber wall temperature at constant $p/T_{wall}$, approximately giving constant gas density in the growth zone. The data points mark the lowest argon gas flow where nanoparticles were observed on the substrate after 10 minutes of deposition. The approximate increase in water vapor pressure as a function of wall temperature is also plotted in the figure (fat black curve). (b) The $Q_{Ar}$ limit as function of oxygen gas flow. The error bars represent the uncertainty of the data point. The different colors represent different experimental days as given in the graphs. The lines between the data points are for eye guidance.

One question to be answered is whether there are trace levels of contaminants in the helium gas that aid in the nucleation process, and whether it is the dilution of these elements at higher argon gas flows that gives rise to the $Q_{Ar}$ limit. To test this, oxygen was intentionally injected into the helium gas and the $Q_{Ar}$ limit was evaluated as a function of the oxygen gas flow. The resulting data can be seen in figure 4 (b) for a pressure of 266 Pa. At oxygen gas flows of $3.75 \times 10^{-3}$ to $5 \times 10^{-3}$ sccm, there is a steep increase in the $Q_{Ar}$ limit, but below $2.5 \times 10^{-3}$ sccm of oxygen, the $Q_{Ar}$ limit was not influenced. This is probably a "getter pump region" where no oxygen reaches the nanoparticle nucleation zone because it gets gettered on the titanium coated vacuum chamber walls. Since trace levels of contaminants inside the He gas supply would be significantly lower than 0.0025 sccm, this experiment shows that contaminations in the helium gas cannot be the source of nucleation seeds.

### 3.3. Nanoparticle structure and size

To evaluate the nanoparticle structure, analysis with transmission electron microscopy (TEM) was performed on nanoparticles synthesized well away from the boundaries of the process window: at 533 Pa, 10 sccm $Q_{Ar}$ and 55 sccm $Q_{He}$. The analysis shows a polydisperse particle size distribution with three distinct diameters at 5, 20 and 50-70 nm. Typical particles within the large size distribution are shown in figure 5. These are weakly faceted, typically with a hexagon, panel (a), or octagonal projection, panel (d), indicating that the particles are crystalline consisting of only one (in this image) or few crystal domains. This is confirmed by high resolution TEM analysis of the core of the particles showing large areas with only a single crystalline orientation. The hexagonal structure is confirmed by high resolution TEM, panel (b) and FFT, panel (c). High angle annular dark-field scanning transmission electron microscopy analysis shows a 3 nm thick, lower atomic mass shell layer surrounding the particles (as evidences by the lower intensity region surrounding the particles in the image of panel (e)). An energy dispersive X-ray spectroscopy scanning TEM (STEM) line scan, panel (f), shows the expected profile for titanium, increasing towards the center of the particle. However, for oxygen the signal is drops slightly across the particle, indicating an oxygen deficient core. The STEM results for the larger particles is consistent with a titanium core surrounded by a 3 nm thick oxide shell [13]. It should be noted that this core-shell structure has not been observed in our previous works with titanium [14] [5] [4], where only titanium monoxide particles were observed at the lowest oxygen flows. It is most probable that this oxide shell is a native oxide shell, *i.e.*, a shell that forms when the particles are removed from the vacuum system and exposed to oxygen in the atmosphere. The formation of a thin oxide layer on metal surfaces exposed to air is well known. It is often named the passivation layer if it inhibits further growth of the oxide. The thickness depends on material, exposure time, humidity, etc. Schultze and Lohrengel [15] give values of the initial oxide thickness after exposure to air for different metals. For Ti they report a thickness of 1.3 to 5.4 nm. The thickness we observe, see Fig. 5(a), is 3-5 nm, thus fitting the expected passivation layer thickness perfectly.

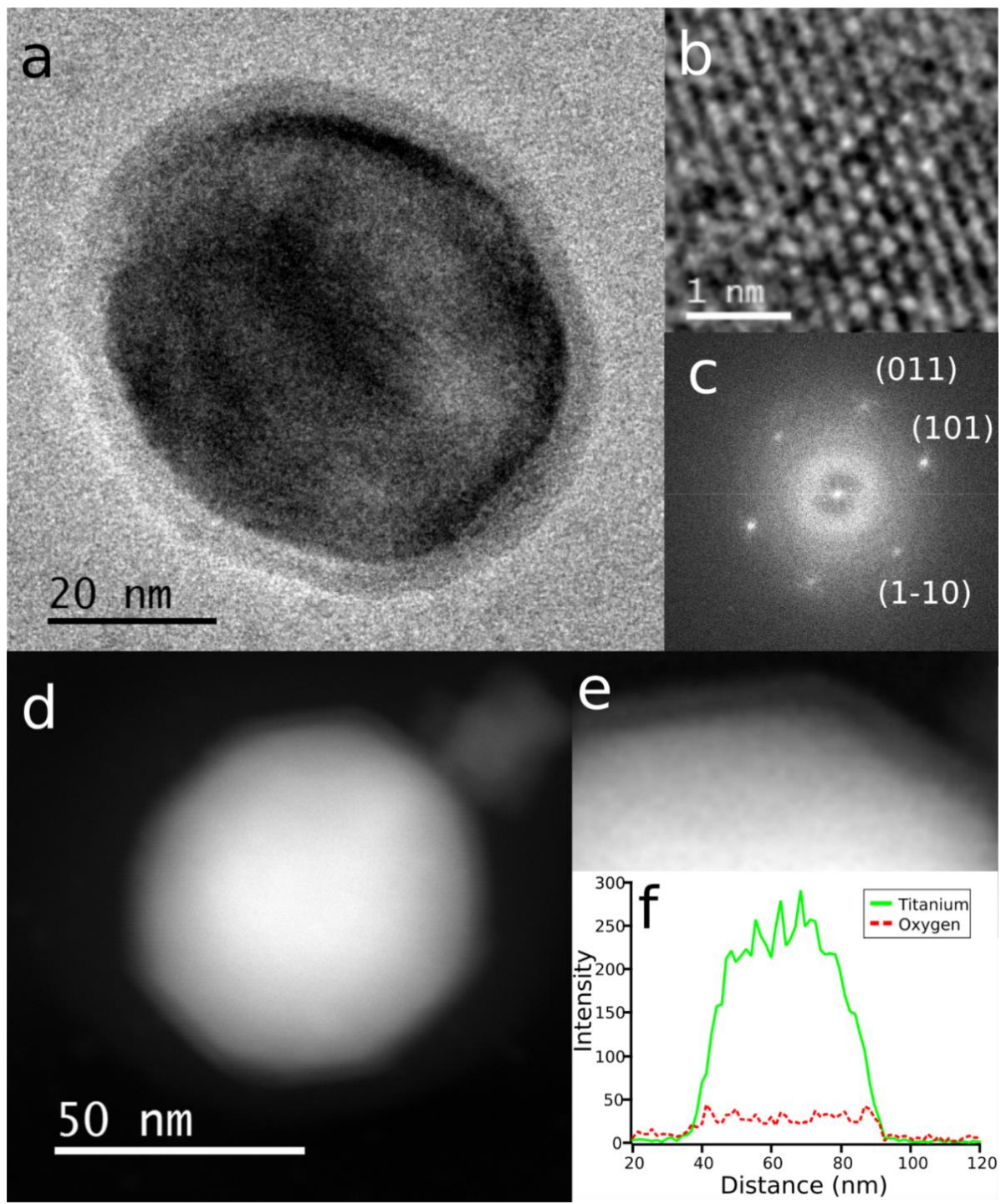

*Figure 5 Nanoparticles synthesized at 533 Pa, 10 sccm Ar and 55 sccm He. (a) BFTEM image of a hexagonal particle, (b) HRTEM image of the core and (c) its associated FFT pattern (indexed to Ti). (d) STEM image of an octagonal particle with (e) zoom in view showing the shell layer and (f) EDS line profiles indicated an oxygen deficient core*

To confirm that the particles were in fact metallic titanium, X-ray diffraction was performed on particles produced at the same process parameter combination: a pressure of 533 Pa, a helium gas flow of 55 sccm, and an argon gas flow of 10 sccm. In figure 6 the titanium (100), (101), (102) and (103) peaks can clearly be seen, which correspond to hexagonal titanium. The other peaks visible in the spectra are from the gold substrate. There were also some unidentifiable peaks around 55°.

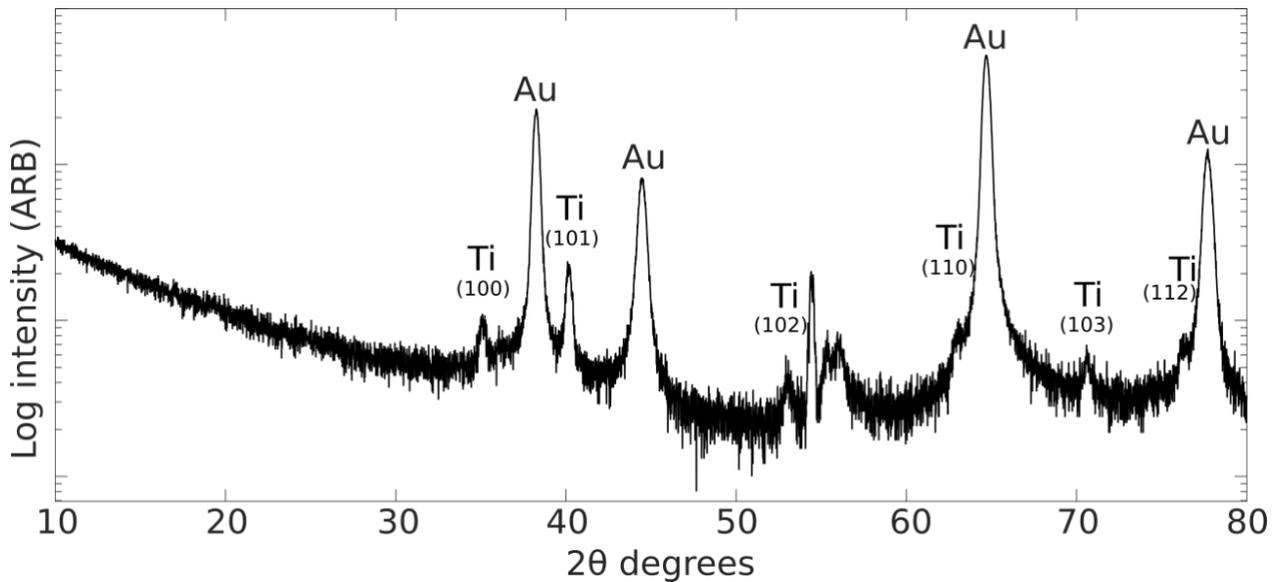

*Figure 6 X-ray diffractograms of nanoparticles synthesized at a process pressure of 533 Pa demonstrating a hexagonal titanium phase.*

The nanoparticle size as a function of pressure was measured and is plotted in figure 7 a. The gas flows were here held constant at $Q_{Ar}$ =10 sccm $Q_{He}$ =55 sccm and the pressure was varied by throttling the pump speed. There is barely any average size increase between 200 and 467 Pa. However, there is a small increase in the width of the size distribution. When the pressure was further increased, there was a steep increase in the nanoparticle size. There are also two distinct size distributions between pressures of 533 to 633 Pa visible in the SEM images. These results are somewhat different from the ones we previously observed in the high vacuum system [5], where the size increased more consistently with the pressure increase. However, in both experiments, higher pressure has a size increasing effect.

As in figure 3, the $p$ limit is drawn as a dotted red line in figure 7 (a). In this case, where a scanning electron microscope was used, the pressure limit was found at 187 Pa. This is close enough to the earlier value of 200 Pa, in figure 3, to validate our ocular inspection *(i.e.,* looking at the substrates with the naked eye) as a good estimate for the limits of getting nanoparticles.

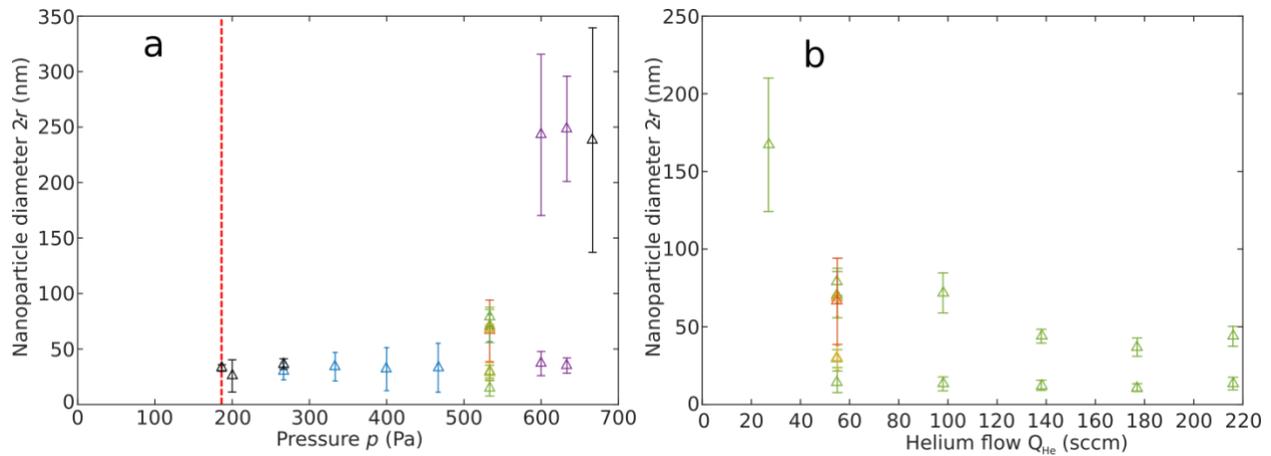

*Figure 7 (a) Nanoparticle size as a function of pressure. The size stays relatively constant until it steeply increases at 533 Pa. Above this pressure there are both small 10-20 nm and large nanoparticles mixed. No particles were found by SEM analysis at pressures below 187 Pa (red dashed line). (b) Nanoparticle size at 533 Pa, as function of helium flow. The nanoparticle size decreases with increasing helium flow.*

The nanoparticle size was also measured as a function of $Q_{He}$, at constant $p$ = 533 Pa and $Q_{Ar}$ =10 sccm, and plotted in figure 7 (b). At all $Q_{He}$ except 27 sccm there were two distinct particle distributions visible. A clear trend can be seen where the larger nanoparticles' size decreases with increasing $Q_{He}$.

The conclusion of the experimental section is that titanium nanoparticles can nucleate and grow, also under clean UHV conditions, but only by the use of a high buffer gas pressure of helium. This nucleation and growth appears to occur without the need for oxygen or water in the process. However, it only occurs within a limited process window, which for our device is shown in figure 3. The pressure has to be above a "$p$ limit" around 200 Pa. When this is satisfied, the argon flow has to be in a range between an "upper $Q_{Ar}$ limit" around 25 sccm, and a lower limit where the discharge becomes unstable.

### 4. Theory

In this section we will explain the arc-like behavior at low $Q_{Ar}$, and also identify a too high gas temperature just outside the hollow cathode as the most probable reason for the disappearance of the particle generation above the upper $Q_{Ar}$ limit. In both cases we will see that the helium gas is involved, although through completely different physical mechanisms.

For this discussion, we divide the growth environment in to 3 zones as shown in figure 8. Zone 1 is inside the hollow cathode. Also between the pulses, the temperature is elevated in this region since the argon gas gets heated by the cathode walls which do not have time to cool down between the pulses. This also means that the gas temperature here is independent of the vacuum chamber wall temperature. The gas is then ejected out into zone 2 where it, in the time between pulses, mixes with the helium gas. The thermal conduction, and the mixing of the gases, leads to a decreased temperature within this zone. In zone 3 it is assumed that the gas has cooled down to the same temperature as the wall temperature, and the helium and argon gas is completely mixed. Zone 2 is the region into which the growth material is ejected during the pulses, and where the densities and the temperature of the species involved in the nucleation process determine whether nanoparticles are formed or not. The sputtered growth material is created in zone 1, and a fraction of it is ejected out into zone 2 where it starts to expand due to diffusion and ambipolar diffusion [16] [5]. This expansion leads to that the growth material density, in zone 2, continuously decreases the closer it is to zone 3. In zone 3, the nanoparticles can continue to grow as long as there is growth material available.

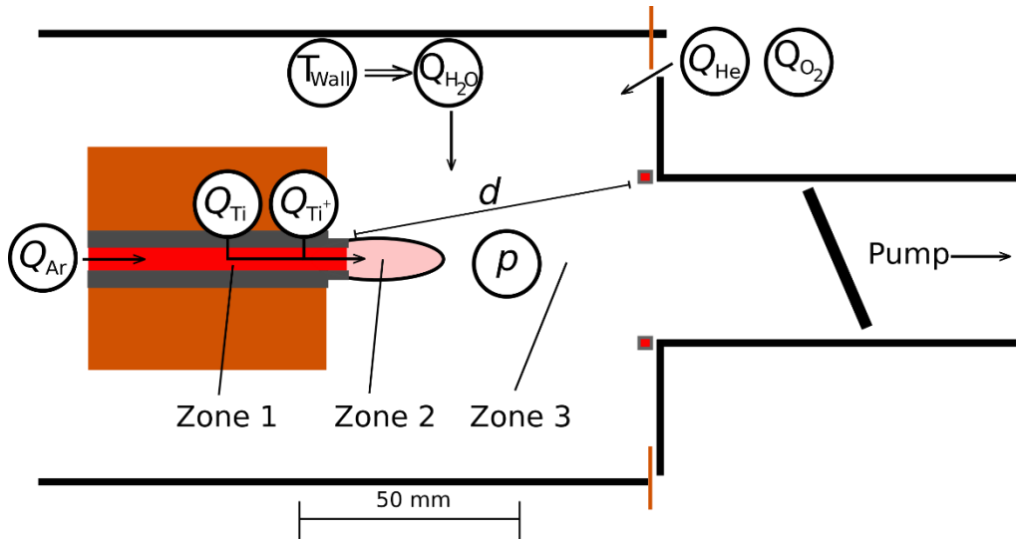

*Figure 8 Overview of the different zones, and of the process parameters which are inscribed in circles. This diagram refers to the situation between pulses. Zone 1 is inside the hollow cathode and has the highest gas temperature, the lowest helium fraction, and the highest growth material (Ti and Ti⁺) density. Zone 2 extends out into the growth zone and here the temperature is elevated relative to the vacuum chamber wall, and the argon gas mixes with the helium gas. In zone 3 the argon gas is assumed to be completely mixed with the helium, and the gas temperature to be the same as the chamber wall temperature. The variable d denotes the shortest distance between the cathode and anode.*

With this model with three zones, we will first explain the process instability limit, observed in figure 2, and then discuss how the nucleation environment, in zone 2, gives rise to the upper $Q_{Ar}$ limit.

### 4.1 Process instability limit

The key to the arcing-like behavior is that the discharge current must pass though zone 3 in order to reach the anode. The effect of increasing the helium flow, for any given total pressure, is the reduction to the density of argon. With a complete mixture of the two process gases assumed in zone 3, the partial pressure of argon $p_{Ar}$ becomes

$$p_{Ar} = \frac{Q_{Ar}}{Q_{He} + Q_{Ar}} p \qquad (1)$$

Solving for $Q_{Ar}$ gives the relation

$$Q_{Ar} = \frac{p_{Ar} Q_{He}}{p - p_{Ar}} \qquad (2)$$

The experimental points for the process instability limit in figure 2 can be well fitted by equation (2) for a constant value of $p_{Ar}$= 28.27 Pa. The blue dot-dashed line shows this curve. Thus, the partial argon gas pressure in zone 3 can be identified as the key to the arc-like behavior. We propose that it is a transition between two types of discharges, from a hollow cathode discharge to a usual glow discharge. In the former case the ionization and the electron confinement *inside* the hollow cathode are important, and a significant part of the discharge voltage falls inside it. In the latter case the discharge can ignite and burn *outside* the hollow cathode, between the outer orifice of the hollow cathode and the anode. The hollow part of the cathode is then not necessary for the discharge. The transition decreases the impedance, and resulting in a higher current for the applied voltage. Since helium has a much higher ionization potential and a much lower mass than argon, it can act as a passive component in the gas mixture, not effectively taking part in the discharge. The condition for forming a glow-type discharge outside the hollow cathode can then be found by looking at the Paschen curve for an argon discharge with a planar cathode and anode. A theoretical curve for the minimum voltage required for a glow discharge breakdown in argon is given by the following equation [17]:

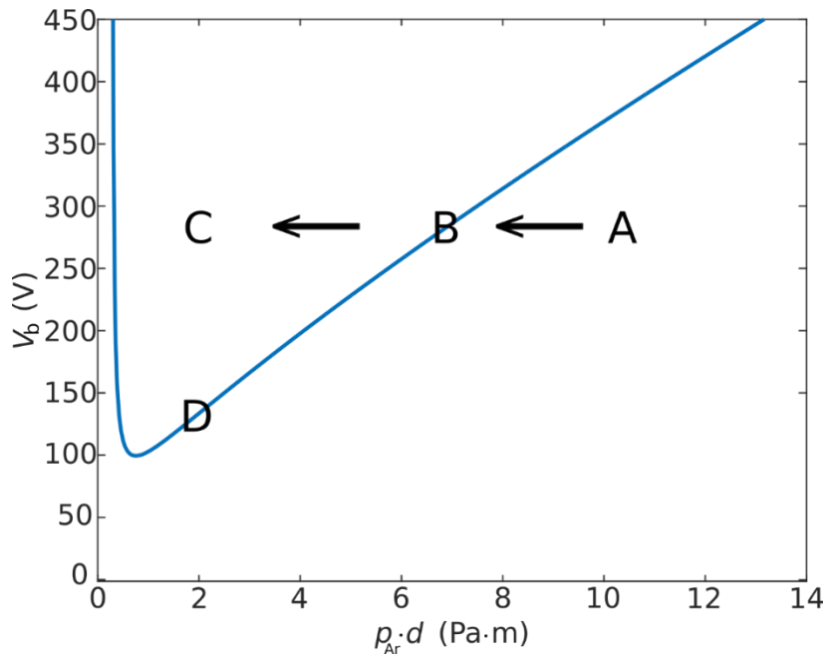

*Figure 9 Paschen curve for the breakdown voltage in pure argon with operational points A, B, C and D drawn for a thought-experiment. A decrease in the argon partial pressure follows the arrows from A to B to C. C corresponds to 1.87 Pascal meter, which is where discharge current suddenly increases in an arc-like fashion.*

$$V_{\text{b}} = \frac{Bpd}{\ln(Apd) - \ln[\ln(1 + 1/\gamma_{\text{s}})]} \quad (3)$$

where $A$ = 8.64 and $B$ = 132 are the gas specific constants for argon [17], $\gamma_{\text{s}}$ is the secondary electron yield which is in the order of 0.1 for $Ar^+$ ion impact on Ti [18], and $d$= 0.066 m is the distance between the cathode and anode, as shown in figure 8. The breakdown voltage is plotted in figure 9 as a function of the pressure times the cathode to anode distance ($p_{Ar} \cdot d$) with operational points A, B, C, and D drawn for a thought-experiment, with a hollow cathode and a HiPIMS power source.

A High Power Impulse Magnetron Sputtering (HiPIMS) source, in contrast to a direct current source, applies the voltage $U_{\text{D}}$ at each pulse start, and can maintain the discharge voltage at high currents. The points A, B, and C correspond to decreasing partial Ar pressure. Let us start with point A which is below the Paschen curve. At this value of $p_{Ar} \cdot d$ a voltage of 375 V is needed to ignite a glow discharge in the growth zone outside the hollow cathode (between the flat ring-shaped end of the cathode and the anode ring). The discharge therefore cannot ignite in a glow discharge mode. Instead it ignites in the hollow cathode mode, where the applied potential drop mainly falls inside the cavity. Outside of the hollow cathode there is only the minimum E field needed to maintain a plasma which can carry the current to the anode. Point B is a transition point, and at point C the discharge becomes unstable since the discharge here can operate either in a hollow cathode mode (where the potential drop falls mainly inside the hollow cathode) or in a glow discharge mode (where the potential drop falls mainly outside). A discharge favors the mode that gives the maximum current for the applied voltage. If it chooses the glow discharge mode it is important to realize that the HiPIMS power source can maintain a voltage above the Paschen curve. Without this ability, the voltage would drop to point D, and the discharge would move into the glow discharge mode. With the voltage maintained at C however, there will instead be a current increase that continues either until it is stopped by the pulse end, until it is interrupted by the arc protection system of the power supply, or until it reaches some stable discharge configuration, for example as an abnormal glow discharge or as an arc discharge.

This identification of the discharge instability mechanism raises possibilities for optimization of the process. For example, the lower boundary of the process window in Fig. 3, the arc-like behavior, should be possible to shift downwards by increasing the anode to cathode distance *d*.

### 4.2. The influences of the process parameters *p*, $T_{\text{wall}}$, $Q_{\text{Ar}}$ on the nucleation process

In this section we will briefly outline the theory for the upper $Q_{\text{Ar}}$ limit. An extended theoretical treatment, including the dimer formation, is given in the companion paper [1].

We are mainly interested in zone 2, where the nucleation is most likely to occur due to the high density of growth material during and just after the pulse. Let us first look at the situation between pulses. Due to the design of the experimental setup, the argon gas passes through the hollow cathode. Since the temperature of the cathode surface is elevated, the gas would obtain a temperature $T_{\text{g}}$ in the order of 1000 K [19] to 1500 K [20] in zone 1. This gas will then cool down in zone 2. The energy to be dissipated is given by:

$$E_{\text{gas}} = mc(T_g - T_{\text{wall}}) \tag{4}$$

where $m$ is the mass and $c$ is the specific heat capacity of the gas. The rate of heat conduction per unit area is given by Fourier´s law:

$$q = -k\nabla T \tag{5}$$

where $\nabla T$ is the temperature gradient across the thermal boundary between zones 2 and 3, and $k$ is the thermal conductivity, which for a gas mixture of helium and argon is given by:

$$k_{\text{mix}} = \sqrt{\frac{k_B^3 T_g}{\pi^3}} \left( \frac{1}{d_{\text{Ar}}^2 \sqrt{m_{\text{Ar}}}\left(1+2.59\frac{X_{\text{He}}}{X_{\text{Ar}}}\right)} + \frac{1}{d_{\text{He}}^2 \sqrt{m_{\text{He}}}\left(1+0.7\frac{X_{\text{Ar}}}{X_{\text{He}}}\right)} \right) \tag{6}$$

where $X_{\text{He}}$ is the mol fraction of helium and $X_{\text{Ar}}$ is the mol fraction of argon [21].

Equation (4) shows that an increased mass flow will increase the amount of gas that has to be cooled down. From equation (6) it becomes evident that the larger the argon gas fraction is, the lower the thermal conductivity becomes and thus there is a lower rate of heat conduction in equation (5). For a numerical example, if $Q_{\text{Ar}}$ is increased from 10 to 20 sccm, the mass to be cooled down increases with a factor of 2 and the thermal conductivity $k_{\text{mix}}$ between the wall and zone 2 decreases by about 20 %. This means that the size of zone 2 increases slightly more than proportionally to the increase in $Q_{\text{Ar}}$. If instead the gas pressure is increased at constant $Q_{\text{Ar}}$, then the velocity of the gas will decrease proportionally. However, on the other hand, the density of the hot gas coming out of the hollow cathode will increase by the same factor. There is thus to a first approximation no net change in the temperature and size of zone 2 for a pressure change.

Now let us estimate the influence of $T_{\text{wall}}$ on the cooling rate of the gas that exits the hollow cathode. $T_{\text{wall}}$ will only influence the cooling rate in zone 2 through the gradient in equation (5). Assuming that the gas exiting the hollow cathode has a temperature of 1250 K, a decrease in $T_{\text{wall}}$ from 425 K to 225 K will only increases the cooling rate by 24 % in the thermal boundary between zones 2 and 3. This is low compared to the influence of $Q_{\text{Ar}}$, and explains why there is, in figure 4 (a), no observable change in the $Q_{\text{Ar}}$ limit at wall temperatures lower than 300 K. This small influence of $T_{\text{wall}}$ thus supports our assumption that the nucleation occurs close to the hollow cathode, in zone 2.

We have now shown that $Q_{\text{Ar}}$ is the dominating parameter for determining the gas temperature distribution in the nucleation zone 2, and therefore propose that the physical reason for the upper $Q_{\text{Ar}}$ limit should involve the gas temperature. We will therefore discuss how the gas temperature influences the initial growth of nanoparticles, with focus on nanoparticles of a size smaller than the thermodynamically stable size $r^*$ (nanoparticles in this size range are sometimes called clusters in the literature, but we chose to use the word nanoparticles independent of size). The nanoparticle temperature ($T_{\text{np}}$) is at least as high as the process gas temperature ($T_{\text{gas}}$) plus a temperature contribution from exothermic reactions on the nanoparticle surface [22]. The heating contribution from these reactions on nanoparticles will be treated in the companion paper [1]. We here only investigate how we can influence the cooling terms. In the free molecular regime as viewed from the perspective of the nanoparticle, the formula for heat transfer from a nanoparticle to the surrounding gas is given by

$$q = \alpha \pi r_{\text{np}}^2 p \sqrt{\frac{2k_B T_{\text{gas}}}{\pi m_g}} \left(\frac{\kappa+1}{\kappa-1}\right)\left(\frac{T_{\text{np}}}{T_{\text{gas}}} - 1\right) \qquad (7)$$

where $r_{\text{np}}$ is the radius of the nanoparticle, $m_g$ is the mass of the gas atom and $\kappa$ is the specific heat ratio [23]. The constant $\alpha$ is the thermal accommodation coefficient, which depends on which type of gas atom that collides with the particle. For collisions with a stainless steel surface, values of $\alpha = 0.866$ for argon and $\alpha = 0.360$ for helium has been measured [24]. If the argon gas is completely substituted by helium, there is only a 31 % increase in the cooling rate of the nanoparticle. Comparing this to the 780 % increase of the thermal conductivity of the gas for the same substitution [25], the cooling rate increase on the nanoparticle is relatively small. We can thus conclude that the primary effect of using helium gas is not to cool the nanoparticle directly but rather indirectly by aiding in the cooling of the hot vapor ejected from the cathode. In the companion paper [1] we will show that it is the gas temperature that is the key internal parameter because even modest gas temperature increases causes sub critical nanoparticles to evaporate and not grow to their thermodynamically stable size. This is because the nanoparticle temperature is directly proportional to the gas temperature [22], and the evaporation rate at temperatures above the activation energy [26] is exponentially dependent on the nanoparticle temperature [27].

## 5. Discussion

The first issue to discuss is whether we can exclude that the nucleation of the nanoparticles is assisted by oxygen, even at low base pressures as in this UHV experimental setup. One way to test this is based on our previous experiments in the high vacuum regime [5], from which we know that an increase in $Q_{\text{Ar}}$, or a decrease in the pressure, would decrease the contaminant oxygen content within the growth zone equally much. If such contamination causes nucleation, a slope with constant $p/Q_{\text{Ar}}$ in the gas flow-pressure 2-d space for the limit of nucleation in the $(p, Q_{\text{Ar}})$ survey would be expected. This type of nucleation limit was also found in [5] and is marked (1) in figure 10 which compares the results from the high vacuum system and the UHV system. The conclusion was that the bottleneck parameter for nucleation in this experiment was a lowest density of contaminants, probably H₂O. If residual contaminants were important for the nucleation also in the UHV system, a similar linear dependence (a line through origo) would result. However, this is clearly not the case for the $Q_{\text{Ar}}$ limit, which is marked (2) in figure 10.

It was, however, possible to replicate the contamination-assisted nucleation also in the UHV system by increasing the degassing rate by heating the chamber wall, see figure 4 (a). The increase in the partial pressure of water, as function of the wall temperature, is drawn in the figure as estimated from the Antonine equation:

$$p_{H_2O} \propto e^{A-\frac{B}{T+C}} \qquad (8)$$

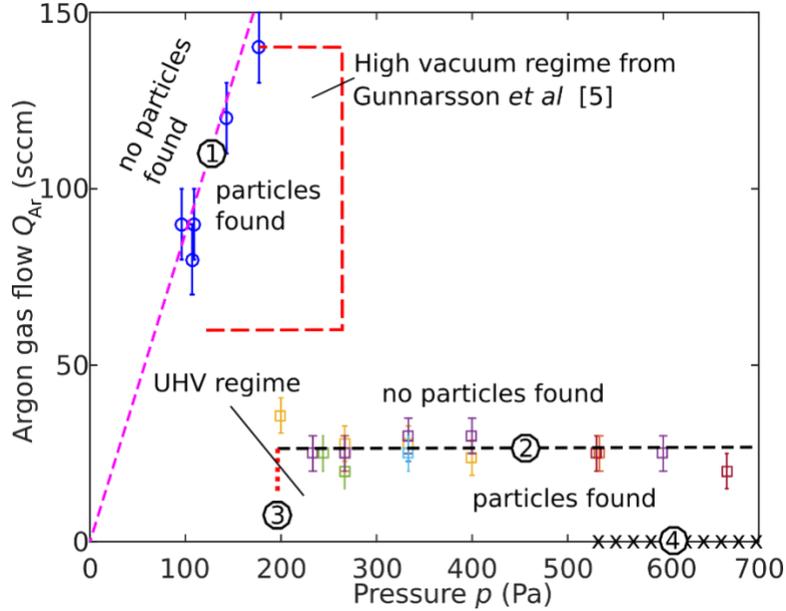

*Figure 10 An overview of the nucleation limits found in $(p, Q_{Ar})$ surveys. Limit no (1) was found earlier in a high vacuum system [5] without helium added to the process. The red dashed area represents the parameter ranges investigated in that experiment. The purple dashed line marked (1) is fitted to the data points (blue circles), and identifies the boundary for when nanoparticles were found. This boundary was proposed to be due to a lower partial pressure of water vapor for combinations of low pressure and high gas flow. There are three corresponding boundaries in this work in the UHV system where helium is added. Nucleation disappears above of the dashed black line, the upper $Q_{Ar}$ limit (2) and left of the dotted red line, the $p$ limit (3). Finally, no nanoparticles were found in pure helium, i. e., without any argon gas flow. This "zero-flow" limit to $Q_{Ar}$ is marked (4).*

where the constants A= 16.39, B =3885.7, and C = -42.15 are specific constants for water [28]. From this equation, we see that the partial pressure has a strong temperature dependence, and would be significantly higher at higher temperatures. In figure 4 (a) we see a clear increase in the $Q_{Ar}$ limit at temperatures higher than 400 K which can be attributed to outgassing according to the Antoine equation. Of most interest here is the expected water vapor pressure at our normal operational wall temperature, 300 K. The black curve in figure 4 (a) shows that it is about two orders of magnitude below that at 400 K. Thus, if the upper $Q_{Ar}$ limit were due to contamination-assisted nucleation, a clear decrease of the $Q_{Ar}$ limit from 400 to 300 K should be seen. However, no measurable change in the slope of the $Q_{Ar}$ limit in figure 4(a) was observed below 400 K, again confirming that contamination-assisted nucleation was not important in the UVH experiment.

To further study when contaminants started to influence the upper $Q_{Ar}$ limit, oxygen was deliberately introduced in the helium gas flow, see figure 4 (b). Interestingly, no clear change in the $Q_{Ar}$ limit could be found when the oxygen gas flow was $Q_{O_2} \leq 2.5 \times 10^{-3}$ sccm. Assuming all gases mixed in zone 3, this would mean an oxygen partial pressure of 8.3·10$^{-3}$ Pa, which is several orders of magnitude higher than what realistically could reach the cathode, since severe cathode poisoning would be expected at so high oxygen partial pressures. Instead the explanation has to be that at these flows, the oxygen reacts with the titanium coated vacuum chamber wall and a too small fraction of it reaches zone 2 to influence the nucleation. This also means that possible trace levels of contaminants in the helium gas would be gettered on the vacuum chamber wall, a third support of the conclusion that the upper $Q_{Ar}$ limit (2) in figure 10 is not determined by some process that involves contamination.

In section 4.2 we proposed that the key parameter for the upper $Q_{Ar}$ limit is the gas temperature $T_g$ in zone 2, mainly through the strong influence of this temperature on the growth from dimers to nanoparticles of a stable radius $r^*$. A higher gas temperature, and thus higher nanoparticle temperature, increases the evaporation rate of sub critical nanoparticles ($r < r^*$), preventing them to grow. A related observation has previously been made by Quesnel *et al* [19] who showed that, in a nanoparticle (cluster) source, there is a narrow region where nucleation of copper nanoparticles is possible. Too close to the cathode, the gas temperature is too high and too far away from the cathode, the density of sputtered material is too low. This explanation is also consistent with the observation that nanoparticles can form at higher $Q_{Ar}$ when oxygen is added to the process, see figure 4 (b). Titanium oxide has lower vapor pressure than titanium [29] [30], and thus titanium oxide nanoparticles will be more stable. As a consequence, they can grow to the stable radius $r^*$, *i.e.* nucleate, at higher gas temperatures (higher $Q_{Ar}$) than pure titanium nanoparticles.

One goal of the present work is to demonstrate that the addition of helium in an UVH system makes possible the growth of pure Ti nanoparticles. This was not possible to achieve in our previously published works in high vacuum systems [4] [5] [14], where a mixture of titanium and oxygen was always present in the particles. There existed no "sweet spot" for the oxygen level in the high vacuum system, where it is high enough to allow nucleation of nanoparticles, but too low for oxide particles to form.

From the TEM analysis of the particles of particles created in the present system (figure 5) it can be seen that they consist of a titanium core and an oxide shell. One important question is how pure the titanium core is. It is difficult to distinguish the EDS peaks from the oxide shell from those in the core of the nanoparticle, and the x-ray diffraction patterns show hexagonal titanium which could have as much as 30 atomic percent of oxide diluted within it if quickly quenched from 1000 K [31]. We can, however, make an estimate of what composition would be expected from the gas densities in the growth zone. Looking in zone 3, when the sputtered titanium atoms and ions have expanded to the radius of the growth tube, they would have a density of 1.7·10$^{17}$ m$^{-3}$ (see the companion paper [1] for the calculations regarding the amount of sputtered material). The base pressure would give a water vapor density in the order of 2.4·10$^{13}$ m$^{-3}$, which probably gets increased by about an order of magnitude due to the baffling of the pump. Assuming that the nanoparticle composition is proportional to the densities of the two gases, we would get only ≈ 0.1 % of oxygen within the particle core. This is of course a very rough approximation that does not take in to account that all the water molecules would not be ionized and that there is a pulse overlap which would influence the titanium density. It also does not take into account that the oxygen is more strongly bound to the titanium, which likely would increase its sticking probability. However, from the large difference between the densities we draw the conclusion that the core consists of metallic hexagonal titanium with a very low oxygen content, covered by a native oxidized shell that forms after the exposure to air.

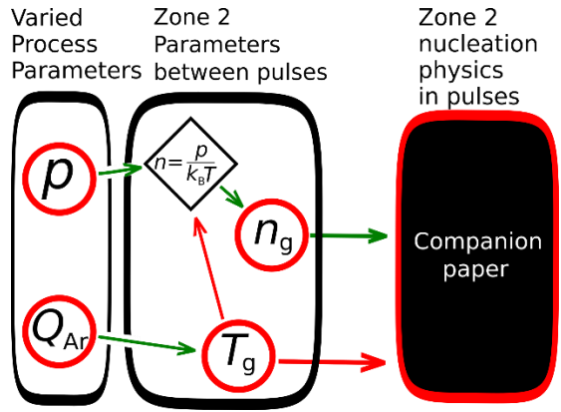

*Figure 11 A reaction flow chart showing how the two varied process parameters in in the $(p, Q_{Ar})$ surveys influence the gas density and temperature in zone 2 between the pulses. Circles denote parameters and diamonds denote processes. Green arrows denote that, an increase in the parameter at the base of the arrow, leads to an increase in the parameter at the arrow tip, and red arrows denote the opposite.*

Both the upper $Q_{Ar}$ limit marked (2) and the pressure limit marked (3) in figure 10 are further investigate in a companion theory paper [1]. To give an interface between the present paper and the theory paper, figure 11 shows a reaction flow chart for the key results from the theory section above which applies to the time between pulses, and in zone 2: a higher pressure *p* directly increases $n_g$ but has little direct effect on $T_g$, and a higher $Q_{Ar}$ increases the temperature directly, but also decreases $n_g$ indirectly: the gas expands at the higher temperature, which gives lower $n_g$. This type of reaction flow chart analysis is continued in the theory paper where it is extended to include the nucleation physics during the pulses.

## 6. Summary

Nanoparticles of hexagonal titanium have been synthesized by pulsed sputtering in a hollow cathode in an ultrahigh vacuum chamber. Introducing helium to the process allows for nucleation to occur without the need for externally supplied oxygen. The process window mapped out was found to be dependent on the argon gas partial pressure, the argon gas flow and the total pressure. At total pressures $p < 533$ Pa, an argon partial pressure of $p_{Ar} > 28$ Pa was required to sustain a stable discharge. Lower $p_\text{Ar}$ resulted in an unstable discharge with an arc-like behavior which limited the process window.

In the stable discharge regime, nanoparticles were found at pressures $p > 200$ Pa and at argon gas flows $Q_\text{Ar} < 25$ sccm. The latter limit is proposed to be due to a temperature increase in the nucleation zone at higher argon gas flows, which increased the evaporation of the growing nanoparticles, preventing them to reach their thermodynamically stable size. This explanation is consistent with the observation that particles can grow at higher argon gas flow if oxygen is supplied to the process, since titanium oxide has a lower vapor pressure compared to titanium. The pressure limit at 200 Pa is discussed in the companion paper [1], and is attributed to a lower dimer formation rate at lower pressures, mainly caused by a faster dilution of the growth material.

The nanoparticle size could be tuned by changing the total pressure and/or the helium gas flow. When removed from the vacuum system, the nanoparticles form a thin oxide shell which makes them stable at ambient conditions. The possibility to synthesize pure titanium nanoparticles without the need of adding oxygen opens up new doors for manufacturing non-contaminated particles from highly reactive materials.

**Acknowledgements**

This work was made possible by financial supported by the Knut and Alice Wallenberg foundation (KAW 2014.0276) and the Swedish Research Council under Grant No. 2008-6572 via the Linköping Linneaus Environment - LiLi-NFM. We also acknowledges financial support from the Swedish Government Strategic Research Area in Materials Science on Functional Materials at Linköping University (Faculty Grant SFO Mat LiU No 2009 00971) and from the Swedish Research Council (VR) Grant No. 2016-05137_4.